\documentclass[preprint,12pt]{elsarticle}

\usepackage{amssymb}
\usepackage{amsmath}

\journal{Extreme Mechanics Letters}

\begin{document}

\begin{frontmatter}



\title{A rediscovery of stiff pentmodes. A comment on "High bulk modulus pentamodes: the three-dimensional metal water"}


\author{Graeme W. Milton} 

\affiliation{organization={Department of Mathematics, The University of Utah},
            addressline={155 South 1400 East, JWB 233}, 
            city={Salt Lake City},
            postcode={84112}, 
            state={Utah},
            country={USA}}

\begin{abstract}
We bring attention to the fact that the claim of  Brambilla et.al.  \cite{Brambilla:2025:HBM}
of discovering a novel design for  pentamode materials is incorrect. Back in 2016 Briane Harutyunyan  and myself  \cite{Milton:2016:PEE} designed a class of stiff pentamodes, that include the high bulk modulus pentamodes of  Brambilla et.al. Our design generalized to three-dimensions, and to full anisotropy, the main aspects of a two-dimensional construction of Sigmund \cite{Sigmund:2000:NCE}. It is emphasized that
the in depth analysis of  Brambilla et.al. goes well beyond our brief treatment. 

\end{abstract}


\begin{highlights}
\item Refutes the claim by Brambilla et.al. \cite{Brambilla:2025:HBM} that they designed a novel class of pentamode materials. 
\item Brings attention to the work of Briane, Harutyunyan  and myself
\cite{Milton:2016:PEE}
who in 2016 had previously discovered this class of pentamode. generalizing a two-dimensional bimode construction of Sigmund \cite{Sigmund:2000:NCE}.
\item Emphasizes that Brambilla et.al. have taken their analysis of the stiff pentamodes much further than our  brief treatment.
\item Remarks that as the stiff pentamode becomes more ideal, the buckling of the thin fibers in each sheaf  becomes an issue if the material is under  compression without prestresses.
\end{highlights}


\end{frontmatter}



\label{sec1}
Pentamodes, as imaged in Figure 1, are a family of materials, introduced in \cite{Milton:1995:WET}  that have drawn considerable attention: see, for example, \cite{Milton:2006:CEM, Norris:2008:ACT, Scandrett:2010:ACU, Cipolla:2011:DIP, Scandrett:2011:BOP, Gokhale:2012:STP, Kadic:2012:PPM, Martin:2012:PBS, Kadic:2013:AVT, Layman:2013:HAE,  Schittny:2013:EMM, Buckmann:2014:EMU, Kadic:2014:PPM,  Cai:2015:PMA, Chen:2015:LPA, Tian:2015:BMA, Amendola:2016:ERA, Fabbrocino:2016:SAP, Huang:2016:PPA, Hedayati:2017:AMM,  Wang:2017:COD, Milton:2018:NOP, Sun:2018:DUA,  Mohammadi:2020:HAP, Li:2021:TDP, Cushing:2022:DCT,Mohammadi:2024:PEU}. 
\begin{figure}[ht!]
	\centering
	\includegraphics[width=0.6\textwidth]{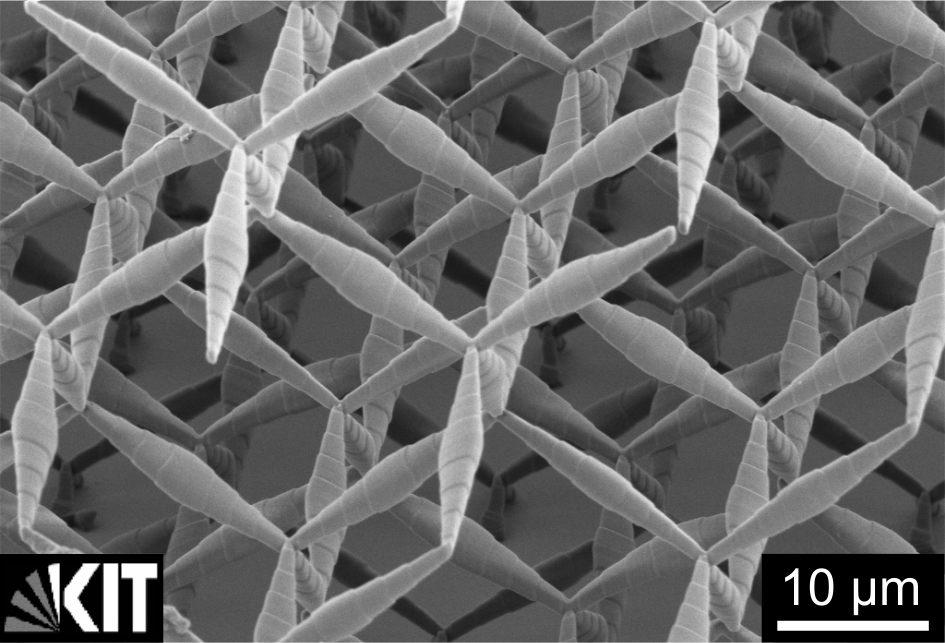}
	\caption{An electron micrograph of the pentamode structure created by Kadic, B\"uckmann, Stenger, Thiel and Wegener \protect\cite{Kadic:2012:PPM} using a 3-dimensional lithography technique. (Used with the kind permission of Martin Wegener.)}
	\label{fig3}
\end{figure}
Like liquids they support only one loading. Unlike liquids or gels, this loading need not be hydrostatic but rather could be a shear loading, or a combination of both hydrostatic and shear loadings. They are compliant to a set of 5 independent loadings (hence the name pentamode). Their elasticity tensor, when represented as a $6 \times 6$ matrix, as is common in mechanical engineering, has 5 small eigenvalues and one much larger eigenvalue. The eigenvector (stress) corresponding to the large eigenvalue can be whatever one likes. 
Ideally, the ratios of the large eigenvalue to the small eigenvalues should be as large as possible. The original pentamode design had stiff double cone elements touching at points on a lattice and surrounded by a very compliant filler material.  Alternatively, the double cone pivot junctions can be joined by narrow necks of the stiff phase and the  filler phase can be void \cite{Kadic:2012:PPM}, thus making their construction more practical.

Pentamodes that are resistant to hydrostatic loadings include gels, pentamodes with double cones
configured a diamond lattice arrangement \cite{Milton:1995:WET}, and the truss structure discovered independently by Sigmund  \cite{Sigmund:1995:TMP} using topology optimization. A liquid is an ideal pentamode resistant to hydrostatic loading, but when placed on a flat surface it flows under gravity rather than keeping a desired shape. Gels have a very small shear modulus so do not flow and have a density close to that of water. Their density can be made to exactly  match that of water, by inserting bubbles to heavier than water inclusions in them.

For some applications, a disadvantage of the early pentamode designs with a given stiff phase is that as the pentamode design becomes more ideal (either by making the filler more and more compliant, or by narrowing the necks joining the double cone elements when they are surrounded by void) is that while the ratio of small to  large eigenvalues  of the elasticity tensor goes to zero, the (comparatively) large eigenvalue also goes to zero. 
One straightforward way to avoid this is to make the stiff phase increasingly stiff, in an appropriate manner, as the filler becomes more compliant or as the necks narrow. In fact it suffices to only stiffen the narrow necks in the structure. 
Brambilla et. al. \cite{Brambilla:2025:HBM} suggest that a more realistic alternative to obtain stiff pentamodes
is to replace the double cones by cylindrical elements,
one  section of each contains a sheaf of thin beams aligned with the cylinder axis. However, their claim that this construction is novel isn't correct. The same class of stiff pentamodes was introduced by  Briane, Harutyunyan and myself in 2016 \cite{Milton:2016:PEE}: see Figure 2. Earlier, as remarked in that paper but overlooked by Brambilla. et.al.,  Sigmund \cite{Sigmund:2000:NCE} had introduced a two-dimensional bimode structure with a hexagonal array of triangular  inclusions connected by an array of  struts: see Figure 3.  Our construction generalizes the relevant aspects of Sigmund's construction to 3-dimensions. It achieves the same sort of 
behavior as our (non-optimal) stiff pentamodes. The primary aspect of Sigmund's construction (not relevant for our purposes)  was to ensures  it is an optimal bimode material, attaining the Hashin-Shtrikman \cite{Hashin:1963:VAT}
bounds on the bulk modulus. Note, however, that while Sigmund extended his structure to three-dimensions (figure 5 in his paper)  it has plates in the design and so has significant shear modulus: it does not have the behavior of a pentamode. 

\begin{figure}[ht!]
	\centering
	\includegraphics[width=1.0\textwidth]{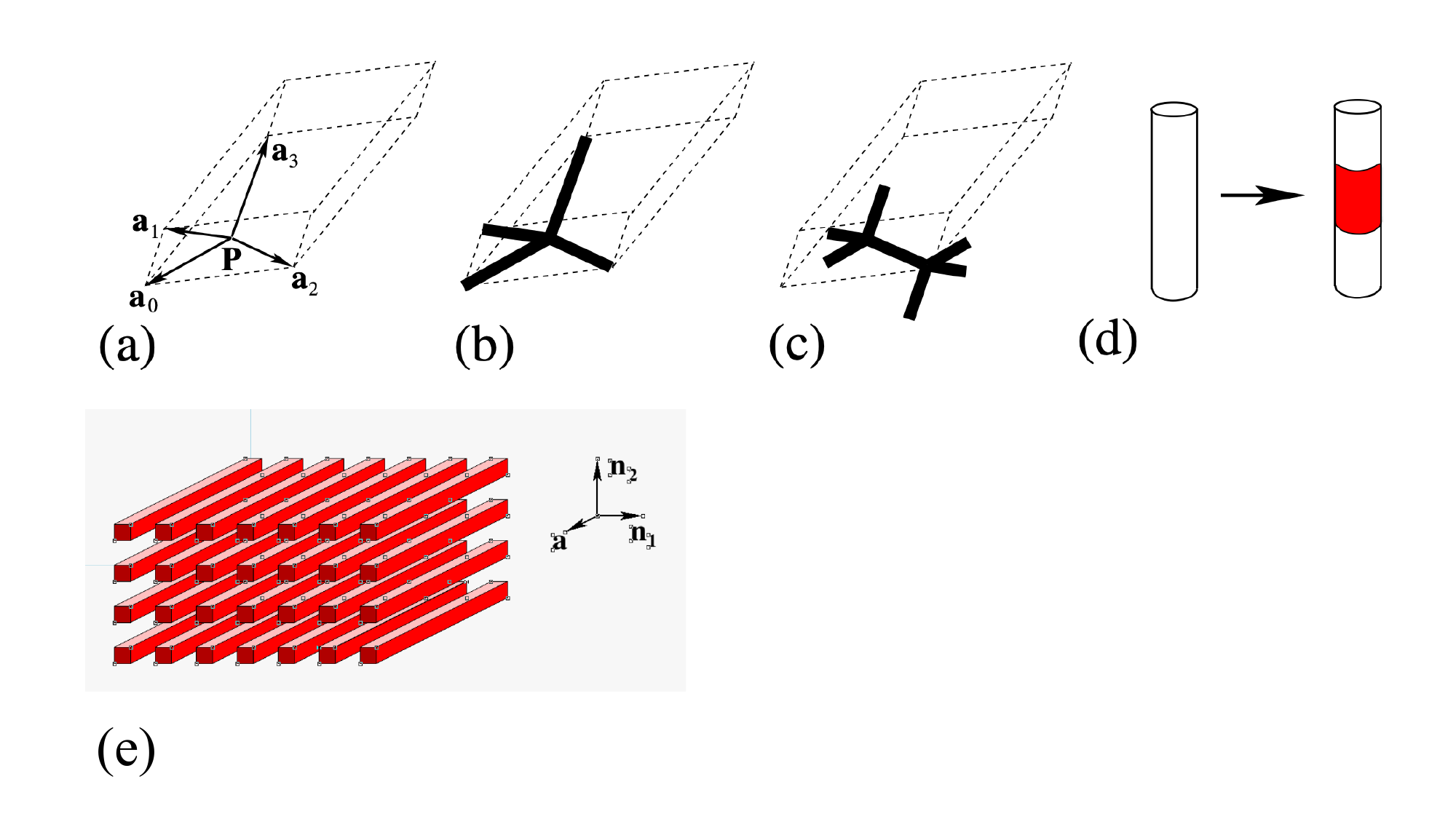}
	\caption{Steps in the construction of stiff pentamodes as presented in \protect{\cite{Milton:2016:PEE}}, figures 9 and 10.  In  (a) the line segments between $P$ and the points $\bf{a_i}, i=0,1,2,3$ are where the double cone structures are placed in the standard pentamode unit cell of periodicity. In (b) these line segments are replaced by cylinders of the stiff phase. (c) shows one such cylinder and two adjacent junctions. In (d) we replace the center section of each cylinder by a sheaf of parallel beams, as in (e), aligned with the cylinder axis.} \label{fig2}
\end{figure}
\begin{figure}[ht!]
	\centering
	\includegraphics[width=0.4\textwidth]{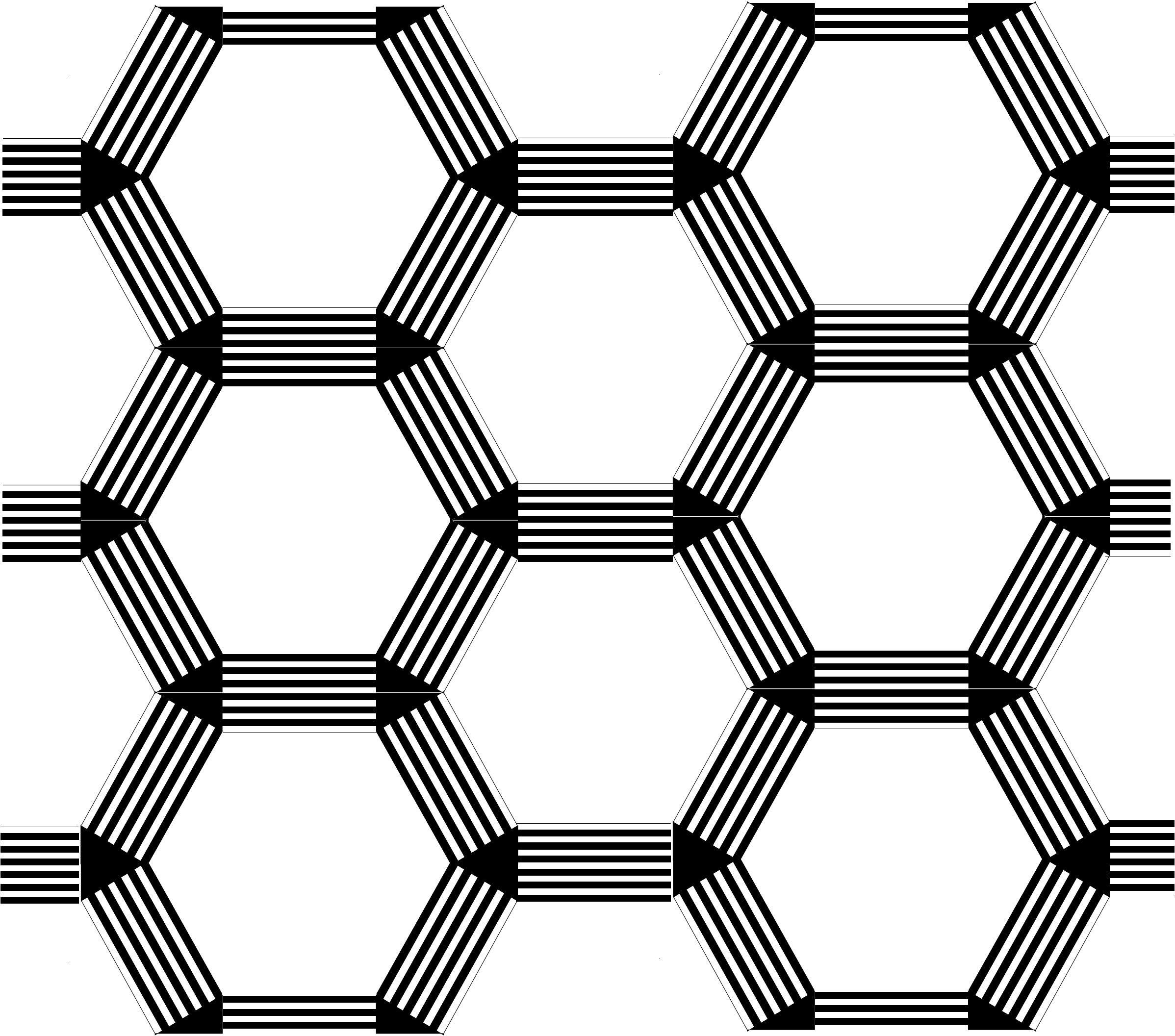}
	\caption{The optimal bimode structure introduced by Sigmund \protect\cite{Sigmund:2000:NCE}. It has the maximal possible effective bulk modulus and an effective shear modulus approaching zero as the elastic moduli of  the phase shown in white approach zero.}
	\label{fig4}
\end{figure}

Our motivation was different to that of Brambilla et.al. They introduced
and carefully analysed  the design as a way to construct three dimensional metal-water, i.e., an elastically isotropic  material with a stiff phase of metal surrounded by air mimicking the bulk modulus of water and having an extremely small shear modulus.  Significantly, they showed that it is not possible in the regular double cone pentamode structure. The mimicking is useful for acoustic cloaking \cite{Norris:2008:ACT, Norris:2011:MWM} as the material has zero acoustic impedance mismatch with water. (The cloak also needs anisotropic pentamodes in the interior to guide stress around the object to be cloaked.) By contrast, our motivation was more theoretical: using them as a stepping stone towards partially characterizing the range of possible elasticity tensors achievable in composites of  given proportions of a chosen isotropic stiff phase surrounded by an arbitrarily compliant filler phase. The stiff pentamodes were inserted as thin layers into materials that optimize sums of elastic energies under different loadings (the so-called Avellenada structures). These thin layers, with stiff pentamode substructures having unit cell size much smaller than the layer thickness, behave as slip planes undergoing infinitesimal slip in prescribed directions under infinitesimal elastic deformations. In particular, one can get a sequence of materials converging to the stiffest possible pentamode material built from the given phases, mixed in the desired volume fractions and supporting any single desired stress. To give the reader insight as to why optimal (and hence stiff) pentamodes are theoretically important we remark, as a side comment, that the optimal pentamodes can be used to get a complete characterization of  the set of all achievable (average stress, average strain) pairs in any composite of these two phases mixed in given proportions \cite{Milton:2018:NOP}.  

It is important to recognize that these constructions are only appropriate 
for linear elasticity: at any finite macroscopic deformation, the pentamode substructures in the thin layers  will undergo large deformations and likely tear so that they no longer function as desired. The same is true, but to a lesser extent, of any of the pentamode
structures. For example, under a finite deformation of pure shear the microstructure will stretch and cause the average stress to be non-zero. By
contrast, there will be no such stress in a fluid as the deformation preserves volume. Additionally, if a stiff pentamode is made increasingly ideal, by making the beams in each sheaf thinner and thinner, then under compression buckling of these beams will be an issue. This can be offset if there are appropriate prestresses. 

Our routes for arriving at the beam substructure within cylindrical elements 
were distinctly different. Brambilla et.al. conceived it by analogy with a rope that is stiff when stretched or compressed, but which is easy to bend. By contrast, we realized that the beam substructure had the desired effective moduli. The idea came from  \cite{Milton:1992:CMP} where it was realized
that  the performance of a particular class of  two-dimensional auxetic materials having a Poisson ratio approaching $-1$ could  be improved by replacing the pivots (similar to the junctions where the double cone elements meet in the standard pentamode)  with a substructure comprised of an sheaf of parallel beams  (i.e., a two-dimensional laminate) \cite{Milton:1992:CMP}: see Figure 3. The linkages between the arms have a uniaxial compressibility proportional to the moduli of the stiff phase but with approximately zero shear resistance (in the ideal limit of the moduli of the compliant phase approaching zero) .  This sort of replacement  is precisely what was needed in our stiff pentamode design and what was done by Brambilla et.al. Thus, the central idea in constructing the stiff pentamodes dates well before the work of Sigmund (who was mainly concerned with finding structures that attain the Hashin-Shtrikman bounds). 

\begin{figure}[ht!]
	\centering
	\includegraphics[width=0.8\textwidth]{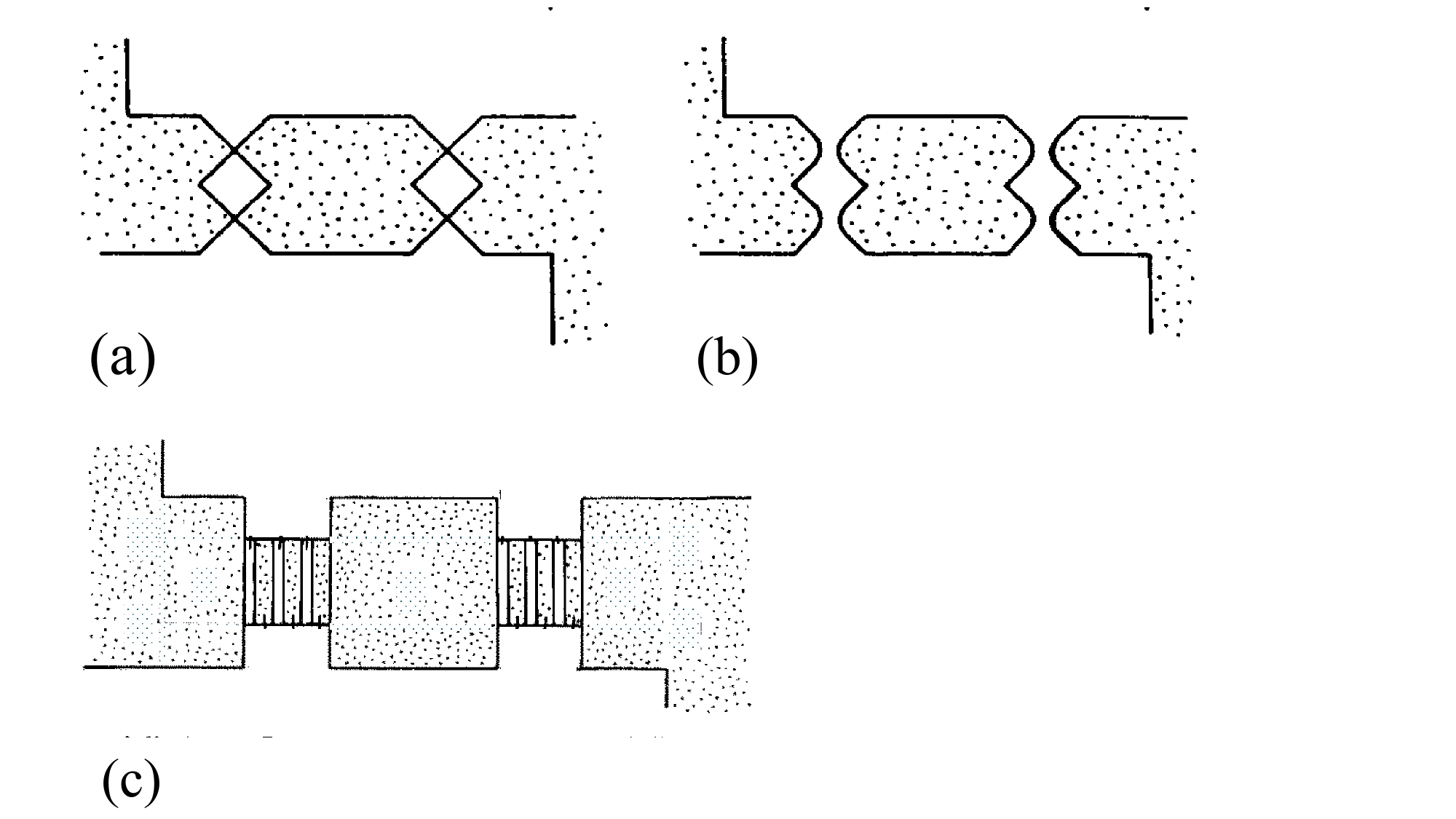}
	\caption{(a) Two linkages that are used to keep the bottom and top surfaces parallel while allowing a sliding motion between them, as needed in the construction of a particular two-dimensional material having Poisson's ratio close to -1, where the stiff phase is white and the highly compliant filler occupies the dotted regions. (b) to make it more realistic one can replace the pivot points in each linkage by thin necks. (c) improved behavior in the limit of high contrast can be obtained by replacing the pivots with a sheaf of beams (a two-dimensional laminate). This improves the stiffness of  each linkage to uniaxial compression while retaining its compliance to shear, exactly what is needed in the stiff pentamode construction. 
	This figure is adapted from  figures 4 and 12 in \protect{\cite{Milton:1992:CMP}}.}
		 \label{fig3}
\end{figure}

By analogy with a rope, Brambilla et.al. also consider cylindrical elements with
a substructure of beams that is then twisted, but they do not analyze this material. As they observe, such a structure has sliding surfaces and quite different responses under stretching and under compression. Therefore, neither their microscopic nor their macroscopic behaviors are that of linear continuum elasticity. 

We emphasize that Brambilla et.al. have taken their analysis of the stiff pentamodes much further than us. Besides presenting beautiful renderings of the structure, they calculate the effective elasticity tensors, mimicking that of water, when Ti 6Al-4V alloy is the stiff phase,  they include an analysis of the effect of each sheaf of thin beams to transmit bending moments, 
they compute the band structure, and they identify the main low frequency modes of deformation.  Notably, to achieve the desired dynamic response, they found it necessary to concentrate the mass at the nodes.


 \section*{Acknowledgments}

The author is grateful to the National Science Foundation for support through Research Grant DMS-2107926.

\ifx \bblindex \undefined \def \bblindex #1{} \fi\ifx \bbljournal \undefined
\def \bbljournal #1{{\em #1}\index{#1@{\em #1}}} \fi\ifx \bblnumber
\undefined \def \bblnumber #1{{\bf #1}} \fi\ifx \bblvolume \undefined \def
\bblvolume #1{{\bf #1}} \fi\ifx \noopsort \undefined \def \noopsort #1{}
\fi\ifx \bblindex \undefined \def \bblindex #1{} \fi\ifx \bbljournal
\undefined \def \bbljournal #1{{\em #1}\index{#1@{\em #1}}} \fi\ifx
\bblnumber \undefined \def \bblnumber #1{{\bf #1}} \fi\ifx \bblvolume
\undefined \def \bblvolume #1{{\bf #1}} \fi\ifx \noopsort \undefined \def
\noopsort #1{} \fi


\end{document}